\documentclass[preprint,authoryear,12pt]{elsarticle}

\usepackage{pdflscape}
\usepackage{hyperref}
\usepackage{amsmath,amssymb,amsfonts}
\usepackage{geometry}

\makeatletter
\def\ps@pprintTitle{%
\let\@oddhead\@empty
\let\@evenhead\@empty
\let\@oddfoot\@empty
\let\@evenfoot\@oddfoot
}
\makeatother

\begin{document}
\begin{frontmatter}
	
	
	
	\title{The Future of Traditional Fuel Vehicles (TFV) and New Energy Vehicles (NEV): Creative Destruction or Co-existence?}
	
	
	\author[1]{Zhaojia Huang}
	\author[2]{Liang Zhang}
	\author[1]{Tianhao Zhi}
	
	\address[1]{Guangdong Provincial Key Laboratory of Interdisciplinary Research and Application for Data Science,
BNU-HKBU United International College, Zhuhai 519087, China}
    \address[2]{Gies College of Business, University of Illinois Urbana-Champaign, USA}
	\cortext[1]{Corresponding author: Tianhao Zhi Email: tianhaozhi@uic.edu.cn}
	
	\newpage
	
	\begin{abstract}
		There is a rapid development and commercialization of new Energy Vehicles (NEV) in recent years. Although traditional fuel vehicles (TFV) still occupy a majority share of the market, it is generally believed that NEV is more efficient, more environmental friendly, and has a greater potential of a Schumpeterian ``creative destruction'' that may lead to a paradigm shift in auto production and consumption. However, less is discussed regarding the potential environmental impact of NEV production and future uncertainty in R\&D bottleneck of NEV technology and innovation. This paper aims to propose a modelling framework based on \citet{Lux1995} that investigates the long-term dynamics of TFV and NEV, along with their associated environmental externality. We argue that environmental and technological policies will play a critical role in determining its future development. It is of vital importance to constantly monitor the potential environmental impact of both sectors and support the R\&D of critical NEV technology, as well as curbing its negative externality in a preemptive manner.
	\end{abstract}
	
	\begin{keyword}
		New Energy Vehicles, Schumpeterian Growth, Externality
		\\
		\emph{JEL classification}: G4, Q27, Q40
	\end{keyword}
	
\end{frontmatter}


\section{Introduction}
\begin{figure}[h!]
	\centering
	\includegraphics[scale=0.8]{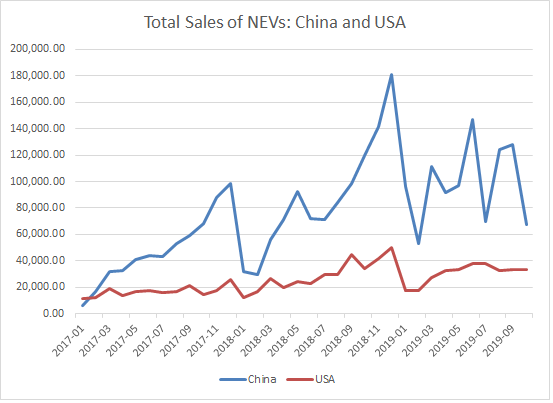}\\
	\caption{\emph{Source:} Wind Database}
	\label{fig:nevs}
\end{figure}

New Energy Vehicles, or NEVs, have become increasingly popular globally in recent years. According to WIND database, the quantity of NEVs had grown from initially about 6,000 to more than 181,000 from 2017 to 2019 in China. The number of NEVs in the US have almost doubled over the same period, as shown in Fig. \ref{fig:nevs}. Although traditional fuel vehicles (TFV) still occupy a majority of the market, it is generally believed that NEV is more efficient, more environmental friendly, and has a greater potential of a Schumpeterian ``creative destruction''\footnote{See \citet{Schumpeter1942}.} that may lead to a paradigm shift in auto production and consumption.

There are numerous research on various factors in NEV consumption and production. \citet{Rezvani2015} and \citet{Gallagher2011} investigate the relationship between consumer preference of NEV in relation to taxation policy and oil price. \citet{Oliver2010} and \citet{Wang2020} found that customers’ environmental awareness contributes to NEV consumption. \citet{Masiero2016} study the role of government subsidy policy in promoting NEV production.

Despite a general optimism in NEV industry at its infant stage, there are at least two sources of uncertainty that may hinder the future of NEV innovation and growth. \emph{First}, it is unclear whether NEV production technology will be constrained by the current technological bottleneck in R\&D in the longer term. \emph{Second}, the production of NEV may also lead to another unforeseen environmental challenge: The production of NEV, especially the permanent magnetic engine demands a high consumption of rare earth elements (REEs) and other pollutants such as lithium. \footnote{Frank LIU, ``EU Says ``No'' to Lithium, and the Underlying Causes Are Worth Attention'', SMM, Retrieved from https://news.metal.com/newscontent/101856562/EU-Says-``No''-to-Lithium-and-the-Underlying-Causes-Are-Worth-Attention/\\Mark Burton. ``Battery Makers Warn EU That Lithium Proposal May Hurt EV Sector'', Bloomberg, Retrieved from https://www.bloomberg.com/news/articles/2022-07-04/battery-makers-warn-eu-that-lithium-proposal-may-hurt-ev-sector
} \citet{Hawkins2013} argues that NEVs have higher environmental burden than ICEVs (Internal Combustion Engine Vehicle) in production. \citet{Adriana2019} found that producing permanent magnets will cause environmental externality. Once reaching a large scale, it may lead to irreversible soil and water pollution that may have severe long-term environmental consequences \citep{Riba2016}. Despite its importance, the potential environmental adversity of NEV production is rarely discussed in current economic literature, which may have a significant potential in providing well-informed economic, as well as environmental policy guidance \citep{Jonathan2000,Adriana2019}.

The aim of this paper is twofold: \emph{first}, we propose a theoretical model in studying the long-term dynamics of TFV and NEV production \& consumption, as well as its environmental externality. We adopt \citet{Lux1995} in modelling the switching and adoption of TFV and NEV from consumer's perspective. \emph{Second}, we discuss the potential policy impact on the long-term dynamics  based on our model from three aspects: (i) \emph{laissez-faire}, (ii) one-sided regulation, (iii) comprehensive regulation.

The rest of this paper is organized as follows: Section \ref{intro} sets up the model. Section \ref{analytical} provides some analytical results of the 2D and 3D model. Section \ref{Policy} discusses the potential impacts of different policy regimes based on numerical simulations of the 4D model. Finally, Section \ref{Conclusion} concludes.

\section{Model Set-up}
\label{intro}
\subsection{Dynamics of transition probability $x$}
\citet{Lux1995} proposes a framework of modelling the heterogeneous interactions of a large quantity of stochastic agents in terms of mean-field interaction. It is assumed that the number of agents is large enough, so that a master equation can be derived that captures the mean-field variable. In economics context, this variable generally implies the average opinion of the economic agents. There are numerous applications that adopt this framework in studying opinion contagion and asset price bubbles \citep{HOMM2005,FRANKE2012,CDGZ2020}.

We adopt \citet{Lux1995} to model the switching mechanism of consumers. Specifically, we assume that there are $2N$ number of consumers that constantly switch between TFV and NEV based on the transitional probability: a high transitional probability implies that the consumers are more likely to switch from one state to another. The transitional probability is determined by various factors that affect agent's decision-making. Here we denote ${N_E}$ as the number of consumers who prefer NEVs while ${N_F}$ who prefer TFVs. To simplify, we define $n={N_E}-{N_F}$ and ${x_t}=n/2N$. Here $x$ has two implications. \emph{First}, $x$ defines the composition of TFV and NEV in the market. $x=1$ implies that all consumers choose NEV, while $x=-1$ implies the opposite. $x=0$ implies that there are equal number of TFV and NEV consumers. Also, as a general case:
\begin{eqnarray}
N_F&=&(1-x)N,
\\
\frac{dN_F}{dt}&=&(1-x)\frac{dN}{dt}-\frac{dx}{dt}N,
\\
N_E&=&(1+x)N
\\
\frac{dN_E}{dt}&=&(1+x)\frac{dN}{dt}+\frac{dx}{dt}N.
\end{eqnarray}

\emph{Second}, $x$ captures the average opinion and herd behaviour of consumers. According to \citet{Lux1995}, the consumers are more likely to switch to one particular state if this is the choice of majority in the hypothetical model.

Here we let $p_t^{-+}$ be the probability per unit of time that a person changes from TFVs to NEVs , and $p_t^{+-}$ be the probability change on an opposite size. According to \citet{Lux1995}, the dynamics of $x$ can be derived as:
\begin{eqnarray}
	\frac{dx}{dt}=p_t^{-+}(1-x_t)-p_t^{+-}(1+x_t).
\label{eq_xlux}
\end{eqnarray}

\citet{Lux1995} points out that the transitional probability possesses three properties: (i) it has to be positive; (ii) the probability should reflect the prevailing opinion (contagion), and (iii) the change of transition probability is in proportion with $x$ ($dp/p = ax$). Hence it can be formulated as:
\begin{eqnarray}
	p_t^{-+}&=&vexp(s),
\label{eq_tp1}
\\
p_t^{+-}&=&vexp(-s),
\label{eq_tp2}
\\s&=&a_0+a_1x+a_2\pi_F+a_3\pi_E,
\label{eq_s}
\end{eqnarray}
where $s$ is the opinion formation index: a higher $s$ implies that the consumers are more likely to switch from TFV to NEV, as \emph{vice versa}. $\pi_F$ and $\pi_E$ are the two environmental externality index caused by TFV and NEV, which will be discussed in details in the following section.

It is important to note that the parameters in Eq. \ref{eq_s} will be our main focus of policy analysis. The constant $a_0$ captures the consumer preference regardless of environmental externality: a positive $a_0$ implies that the consumers prefer NEV than TFV. This could be due to the technological \& economic advantage of NEV than TFV, as well as from the government subsidy that renders NEV economically more attractive. $a_1$ is the herding parameter: a larger $a_1$ implies that consumers are more prone to follow decision of the majority: if a majority of the consumers choose NEV, then it is more likely that the remaining TFV consumers will switch to NEV, and \emph{vice versa}. The parameter $a_2$ and $a_3$ are our regulatory levers. As the government tries to manage carbon emission in various types of policies such as carbon tax, TFV becomes less attractive, prompting more consumers to switch to NEV. Therefore, $a_2 > 0$. Similarly, if the government strives to regulate NEV pollution, it will make NEV less attractive, thus $a_3 < 0$.

Substituting Eq. \ref{eq_tp1}-\ref{eq_tp2} into Eq. \ref{eq_xlux}, we obtain:
\begin{eqnarray}
	\frac{dx}{dt}&=&v[(1-x_t)exp(s)-(1+x_t)exp(-s)],
\\
s&=&a_0+a_1x+a_2\pi_F+a_3\pi_E.
\end{eqnarray}

\subsection{Dynamics of Environmental Externality $\pi$}
We define two types of environmental externality: \emph{(i)} the instant externality from production process (\emph{Type I}), and \emph{(ii)} the ongoing externality that stems from usage (\emph{Type II}). Both TFVs and NEVs would potentially produce both types of externality from production and usage. Yet for the sake of analytical simplicity, we assume in this paper that NEV is the primary contributor of \emph{Type I }externality (soil and water pollution due to REE), and TFV is the primary contributor of \emph{Type II} externality (carbon emission)\footnote{In reality however, NEV can also contribute significantly to \emph{type II} externality. In the context of electric car for example, it depends on how electricity is generated. Simiarly, TFV production also contributes to \emph{Type I} externality. The assumption here is simply for the sake of analytical simplicity.}. Suppose $N_x$ is the number of vehicles, then \emph{Type I} externality ought to be in proportion with the increase of $N_x$ ($dN_x$), and \emph{Type II} externality in proportion with $N_x$ itself. We assign two parameters $\theta_x$ and $\gamma_x$ as these two types of externality parameters respectively. Furthermore, the ecological system exhibits self-purification, implying that in the long-term, the environmental pollution will reduce from its own. We assign the two parameters $\alpha_1$ and $\alpha_2$ captures the self-purification adjustment rate.
\begin{eqnarray}
	\frac{d\pi_F}{dt} &=& \gamma_F N_F - \alpha_1 \pi_F,
	\\
	&=& \gamma_F N(1-x) - \alpha_1 \pi_F,
	\\
	\frac{d\pi_E}{dt} &=& \theta_E \frac{dN_E}{dt} - \alpha_2 \pi_E,
	\\
	&=& \theta_E [\frac{dN}{dt} (1+x) + \frac{dx}{dt} N] - \alpha_2 \pi_E.
\end{eqnarray}

\subsection{3D System of $\pi_F$, $\pi_E$ and $x$}
Based on previous set-up, we propose the 3D system of $x$, $\pi_F$, and $\pi_E$:
\begin{eqnarray}
	\frac{dx}{dt} &=& v[(1-x)exp(s)-(1+x)exp(-s)],
\label{eqn_x}
	\\
	\frac{d\pi_F}{dt} &=& \gamma_F N(1-x) - \alpha_1 \pi_F,
\label{eqn_pif}
	\\
	\frac{d\pi_E}{dt} &=& \theta_E [\frac{dN}{dt} (1+x) + \frac{dx}{dt} N] - \alpha_2 \pi_E,
\label{eqn_pie}
\end{eqnarray}
where:
\begin{eqnarray}
	s &=& a_0 + a_1 x + a_2 \pi_F + a_3 \pi_E.
\end{eqnarray}

If the total number of vehicle $2N$ is fixed, then $\frac{dN}{dt}=0$. Hence:
\begin{eqnarray}
	\frac{dx}{dt} &=& v[(1-x)exp(s)-(1+x)exp(-s)],
\label{eqn_x}
	\\
	\frac{d\pi_F}{dt} &=& \gamma_F N(1-x) - \alpha_1 \pi_F,
\label{eqn_pif}
	\\
	\frac{d\pi_E}{dt} &=& \theta_E \frac{dx}{dt} N - \alpha_2 \pi_E,
\label{eqn_pie}
\end{eqnarray}
where:
\begin{eqnarray}
	s &=& a_0 + a_1 x + a_2 \pi_F + a_3 \pi_E.
\end{eqnarray}

This 3D system becomes our baseline model of analysis. It is possible to manipulate the policy as well as environmental parameters to analyse different policy regimes. This will be discussed further in the following sections.

\subsection{The 4D Model: The long-term growth rate of N ($g_N$)}
The baseline model discussed in previous section does not allow growth in total number of cars. In reality, $N$ may grow exponentially over time. Here we incorporate the exponential growth of $N$ into the system, which yields:
\begin{eqnarray}
	\frac{dx}{dt} &=& v[(1-x)exp(s)-(1+x)exp(-s)],
\label{eqn_x}
	\\
	\frac{d\pi_F}{dt} &=& \gamma_F N(1-x) - \alpha_1 \pi_F,
\label{eqn_pif}
	\\
	\frac{d\pi_E}{dt} &=& \theta_E [\frac{dN}{dt} (1+x) + \frac{dx}{dt} N] - \alpha_2 \pi_E,
\label{eqn_pie}
\\
\frac{dN}{dt} &=& g_N N,
\end{eqnarray}
where:
\begin{eqnarray}
	s &=& a_0 + a_1 x + a_2 \pi_F + a_3 \pi_E.
\end{eqnarray}

The next two sections will discuss the analytical side of the model, as well as numerical analysis \& policy experiments. We use the 3D model for analytical discussion for the sake of mathematical simplicity, and the 4D model on numerical simulation.
\section{Local Stability Analysis}
\label{analytical}
\subsection{Analysis of the two-dimensional system}
This section analyse the 3D model in terms of its fixed points and local stability. We first reduce the 3D model into 2D model by assuming no rare earth pollution has existed yet, which implies that $\theta_E = 0$ and $\pi_E = 0$:
\begin{eqnarray}
	\frac{dx}{dt} &=& v[(1-x)exp(s)-(1+x)exp(-s)],
\label{x_2d}
	\\
	&=&2v[tanh(s)-x]cosh(s),
	\\
	\frac{d\pi_F}{dt} &=& \gamma_F N(1-x) - \alpha_1 \pi_F,
\label{pif_2d}
\end{eqnarray}

where:
\begin{eqnarray}
	s &=& a_0 + a_1 x + a_2 \pi_F.
\end{eqnarray}

We set $LHS=0$ on both equations and derive the following isoclines:
\begin{eqnarray}
	0 &=& 2v[tanh(s)-x]cosh(s),
	\\
	0 &=& \gamma_F N(1-x) - \alpha_1 \pi_F.
\end{eqnarray}

We first analyse the steady state an local stability of the above 2D system. The Jacobian is derived as:
\begin{eqnarray}
	J &=& \begin{bmatrix}
		F_{x} & F_{\pi_F}\\
		G_{x} & G_{\pi_F}\\
	\end{bmatrix}
	\\
	&=& \begin{bmatrix}
		2v(a_1cosh(s)-cosh(s)-a_1xsinh(s)) & 2va_2(cosh(s)-x sinh(s))\\-\gamma_F N &-\alpha_1
	\end{bmatrix}.
\end{eqnarray}

Since it is difficult to find the exact solution of the fixed point in this equation. As a special case, we set $a_0=-\frac {\gamma_F N}{\alpha_2}$ and $a_2=1$, and we can easily obtain an equilibrium ($x^\star=0$, $\pi_F=-\frac {\gamma_F N}{\alpha_2}$). Furthermore, there is potentially an emergence of other equilibria, depending on the value of economic factor $a_0$ and $CO_2$ emission psychological account constant $a_2$ if and only if $a_0+a_2\pi_F=0$. Therefore, the trace and determinant of the Jacobian at a equilibrium (x = 0) is calculated as:
\begin{eqnarray}
 Det(J)&=& -2v\alpha_1(a_1-1)+2va_2\gamma_F N,
	\\
Tr(J) &=& -\alpha_1+2v(a_1-1).
\end{eqnarray}
The necessary and sufficient condition for the local stability for this equilibrium is that $Det(J)>0$ and $Tr(J)<0$. This condition can be rewritten as: $\alpha_1(a_1-1) > a_2\gamma_F N$, $\alpha_1 > 2v(a_1-1)$. This condition implies that the self-purification parameter $\alpha_1$ plays a crucial role in maintaining the stability of this special fixed point.

\subsection{Analysis of the three-dimensional system}
In this section, we introduce the production pollution $\pi_E$ in our model. We obtain the system as follows:
\begin{eqnarray}
	\frac{dx}{dt} &=& v[(1-x)exp(s)-(1+x)exp(-s)],
	\\
	&=&2v[tanh(s)-x]cosh(s),
	\\
	\frac{d\pi_F}{dt} &=& \gamma_F N(1-x) - \alpha_1 \pi_F,
	\\
	\frac{d\pi_E}{dt} &=& \theta_E \frac{dx}{dt} N - \alpha_2 \pi_E,
\end{eqnarray}
where $\theta_E >0$ and $\alpha_2>0$ are  constants and
\begin{eqnarray}
	s&=&a_0+a_1x+a_2\pi_F+a_3\pi_E.
\end{eqnarray}

The Jacobian of this 3D system is
\begin{eqnarray}
	J=
	\begin{bmatrix}
		F_{xx}&F_{\pi_F x}&F_{\pi_E x}
		\\
		F_{\pi_F x}&F_{\pi_F\pi_F}&F_{\pi_F\pi_E}
		\\
		F_{\pi_E x}&F_{\pi_E \pi_F}&F_{\pi_E \pi_E}
	\end{bmatrix}.
\end{eqnarray}
where
\begin{eqnarray}
	F_{xx}&=&2v(a_1cosh(s)-cosh(s)-a_1xsinh(s)),\\
	F_{\pi_F x}&=&2va_2(cosh(s)-xsinh(s)),\\
	F_{\pi_E x}&=&2va_3(cosh(s)-xsinh(s)),\\
	F_{\pi_F x}&=&-\gamma_F N,\\
	F_{\pi_F\pi_F}&=&-\alpha_1,\\
	F_{\pi_F\pi_E}&=&0,\\
	F_{\pi_E x}&=&2\theta_ENv(a_1cosh(s)-cosh(s)-a_1xsinh(s)),\\
	F_{\pi_E \pi_F}&=&\theta_EN(2va_2(cosh(s)-xsinh(s))),\\
	F_{\pi_E \pi_E}&=&-\alpha_2+\theta_EN(2va_3(cosh(s)-xsin(s))).
\end{eqnarray}
We set
\begin{eqnarray}
	0&=&2v[tanh(s)-x]cosh(s),
	\\
	0&=& \gamma_F N(1-x) - \alpha_1 \pi_F,
	\\
	0&=& \theta_E \frac{dx}{dt} N - \alpha_2 \pi_E.
\end{eqnarray}

Since the close-form solution for the fixed point is hard to obtain, we discuss a special case with $a_0=-\frac{\gamma_FN}{\alpha_1}$,$a_2=1$ and $a_3=-1$, so that we have a mid-point equilibrium for $x = 0$. In this special case, the fixed point is derived as: $x*=0$, $\pi_F=\frac {\gamma_F N}{\alpha_1}$, $\pi_E=0$. The Jacobian is given by:
\begin{eqnarray}
	J&=&
	\begin{bmatrix}
		2v(a_1-1)&2v&-2v
		\\
	-\gamma_FN&-\alpha_1&0
		\\
		2v\theta_EN(a_1-1)&2vN\theta_E&-\alpha_2-2vN\theta_E
	\end{bmatrix}.
\end{eqnarray}
which has the sign structure
\begin{eqnarray}
	\begin{bmatrix}
		?&+&-
		\\
		-&-&0
		\\
		?&+&-
	\end{bmatrix}.
\end{eqnarray}
In a 3D dynamical system, the Routh-Hurwitz necessary and sufficient condition for local stability is:
\begin{eqnarray}
	Det(J)&<&0,
	\\
	Tr(J)&<&0,
	\\
	J_1+J_2+J_3&>&0,
	\\
	-Tr(J)(J_1+J_2+J_3)+|J|&>&0,
\end{eqnarray}
where
\begin{eqnarray}
	J_1&=&
	\begin{vmatrix}
		-\alpha_1&0
		\\
		2vN\theta_E&-\alpha_2-2vN\theta_E
	\end{vmatrix},
	\\
	J_2&=&
	\begin{vmatrix}
		2v(a_1-1)&-2v
		\\
		2v\theta_EN(a_1-1)&-\alpha_2-2vN\theta_E
	\end{vmatrix},
	\\
	J_3&=&
	\begin{vmatrix}
		2v(a_1-1)&2v
		\\
		-\gamma_FN&-\alpha_1
	\end{vmatrix}.
\end{eqnarray}

The trace and determinant of the Jacobian at this fixed point is calculated as :
\begin{eqnarray}
	Det(J)&=& (-\alpha_2-2vN\theta_E)(-2v\alpha_1(a_1-1)+2\gamma_FNv)\\
	&&-2v(-2\gamma_FN^2\theta_Ev\theta_E+\alpha_1\theta_E(2vN(a_1-1))),
	\\
	Tr(J) &=& -2v\alpha_1(a_1-1)+2v\gamma_F N-\alpha_2-2vN\theta_E.
\end{eqnarray}
The necessary and sufficient condition of local stability for this fixed point follows Routh-Hourwitz theorem.

\section{Simulation \& Policy Analysis}
\label{Policy}
This section provides some numerical simulation of the 3D model, which aims to bring some insights of how different policy measures would affect the long-term dynamics of TFVs and NEVs, as well as its associated environmental externality of both types.
\subsection{Scenario 1: Laissez-Faire}
We first simulate an initial situation where the government does not interfere. Also, we assume that there is no difference between TFV and NEV in terms of constant consumer preference ($a_0 = 0$) as an initial condition. In other words, the consumer will is indifferent between TFV and NEV in the absence of government intervention. Neither side of the environmental externality ($\pi_n$ and $\pi_f$) is regulated ($a_2=a_3=0$). The growth rate of total number of vehicles are assumed to be zero ($g_n = 0$) as an initial setting. We first look at the corresponding Jacobian of scenario 1 before running the simulation:
\begin{eqnarray}
	J^{(1)}=
	\begin{bmatrix}
		2v[a_1cosh(a_1x)-cosh(a_1x)-a_1xsinh(a_1x)]&0&0
		\\
		-\gamma_FN&-\alpha_1&0
		\\
		0.2N(2v[a_1cosh(a_1x)-cosh(a_1x)-a_1xsinh(a_1x)])&0&-\alpha_2
	\end{bmatrix}.\label{Eq1}
\end{eqnarray}

The trace and determinant of the Jacobian under this scenario is calculated as:
\begin{eqnarray}
	Det(J) &=& 2\alpha_1\alpha_2v[a_1cosh(a_1x)-cosh(a_1x)-a_1xsinh(a_1x)],
	\\
	Tr(J)& =& 2v[a_1cosh(a_1x)-cosh(a_1x)-a_1xsinh(a_1x)]-\alpha_1-\alpha_2.
\end{eqnarray}

Besides, $J^{(1)}_1=\alpha_1\alpha_2, J^{(1)}_2=-2\alpha_2v[a_1cosh(a_1x)-cosh(a_1x)-a_1xsinh(a_1x)], J^{(1)}_3=-2\alpha_1v[a_1cosh(a_1x)-cosh(a_1x)-a_1xsinh(a_1x)]$.

\begin{figure}[h]
	\centering
	\includegraphics[scale=0.8]{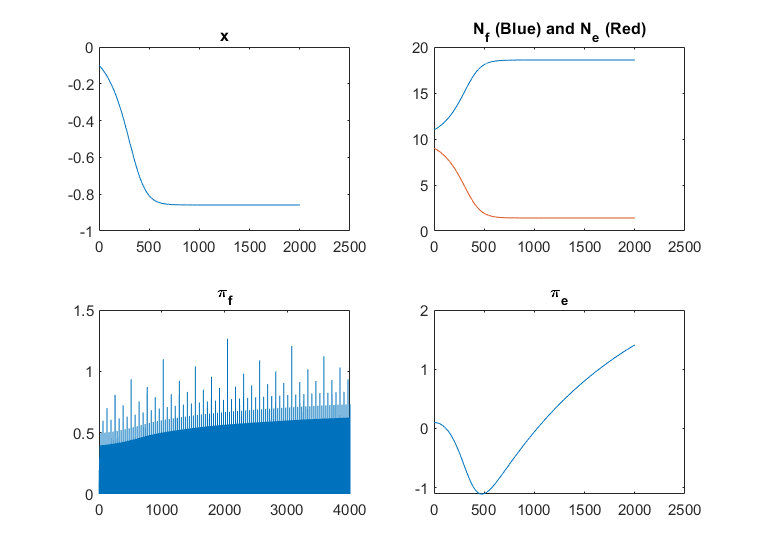}\\
	\caption{Initial Run with $a_1 =1.5$, \emph{ceteris paribus}}
	\label{fig:sim_1_1}
\end{figure}

\begin{figure}[h]
	\centering
	\includegraphics[scale=0.8]{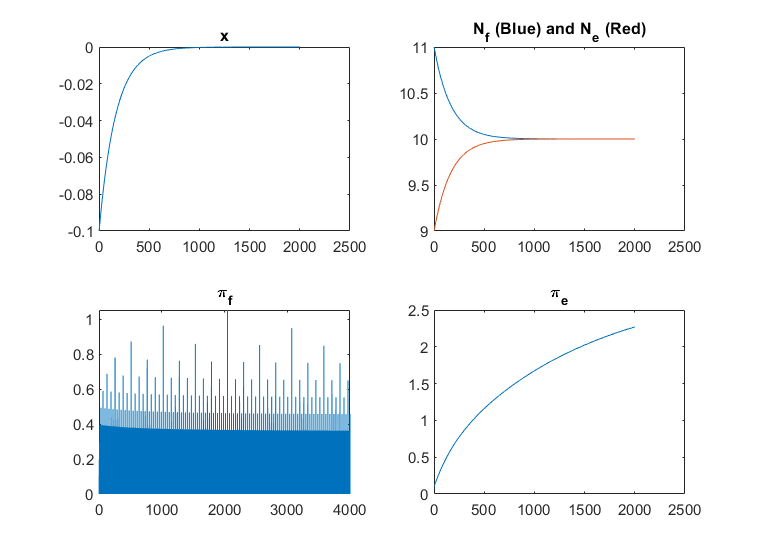}\\
	\caption{Initial Run with $a_1 =0.5$, \emph{ceteris paribus}}
	\label{fig:sim_1_2}
\end{figure}

In this scenario, the herding parameter plays a crucial role, since the main driving force behind the model is the \citet{Lux1995} dynamics. We also look at two sub-scenarios where the herding behaviour is relatively weak ($a_1 = 0.5$) and strong ($a_1 = 1.5$). In this situation, the initial condition of x will play a crucial role in determining the dynamics of the system. Other parameters are set as follows: $N = 10$, $\gamma_f = 0.9$, $\theta_n = 0.2$, $v = 0.6$, $\alpha_1 = 0.03$, $\alpha_2 = 0.07$. The simulation results are provided below:

As shown in Fig. \ref{fig:sim_1_1}, with a strong herding parameter ($a_1 = 1.5$) and an initial advantage of TFV ($x_0=-0.1$), the system eventually converges to a situation where a majority of the consumers opt for TFVs, and both $\pi_f$ and $\pi_e$ rose steadily over the period. On the other hand, when the herding parameter is relatively weak ($a_1 = 0.5$), the number of TFV and NEV eventually becomes equal, \emph{ceteris paribus}. This is shown in Fig. \ref{fig:sim_1_2}

It is also worth to mention that in this scenario, the parameter $a_0$ plays a crucial role in determining whether the system will eventually converge toward TFV or NEV equilibrium in the absence of government interference ($a_0$ is solely determined by economic utility). The previous simulation assumes that $a_0 = 0$, implying that NEV and TFV have the same economic utility to consumers. Given a strong herding parameter and an initial state of TFV majority, the system converges toward an TFV equilibrium, as the simulation shows. However, if NEV is economically more advantageous ($a_0 > 0$), final equilibrium will favour NEV, and \emph{vice versa}.

\subsection{Scenario 2: One-sided Regulation}
The second scenario looks at a situation when the government imposes one-sided regulation, in which the government subsidizes NEV and punishes TFV, overlooking the potential environmental externality caused by NEV production. The support not only makes NEV more economically attractive through tax benefit and other means ($a_0=1$), but also ignores the potential environmental externality caused by NEV production at its initial phase ($a_3=0$). The rest of the parameters are set the same as in scenario 1 discussed previously, with $a_1 = 1.5$ implying a relative strong herd incentive in consumers in a realistic sense. Also, we assume that $N$ grows exponentially at $10\%$ \emph{per annum} over time ($g_N=0.1$).

\begin{figure}[h]
	\centering
	\includegraphics[scale=0.8]{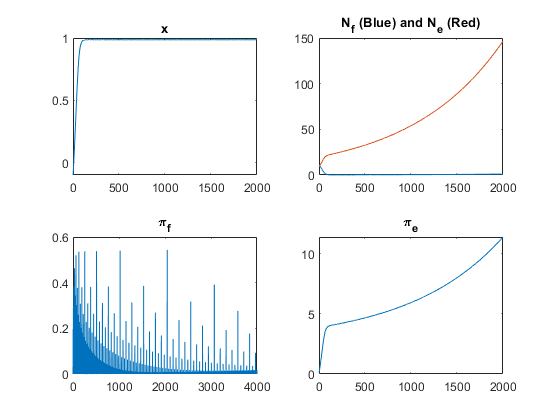}\\
	\caption{Scenario 2 (\emph{``Creative Destruction''})}
	\label{fig:sim_2}
\end{figure}

As shown in Fig. \ref{fig:sim_2}, in such scenario, most consumers almost instantly choose NEV due to the unparallel advantage from the one-sided policy, and the carbon-related externality from TFV diminishes over time. However, there is a drastic increase of $\pi_e$ arising from the close-to-exponential growth of NEVs. The long-term environmental impact of NEV production can not be ignored, once the NEV production reaches a large scale.

\subsection{Scenario 3: Comprehensive Regulation}
\begin{figure}[h]
	\centering
	\includegraphics[scale=0.8]{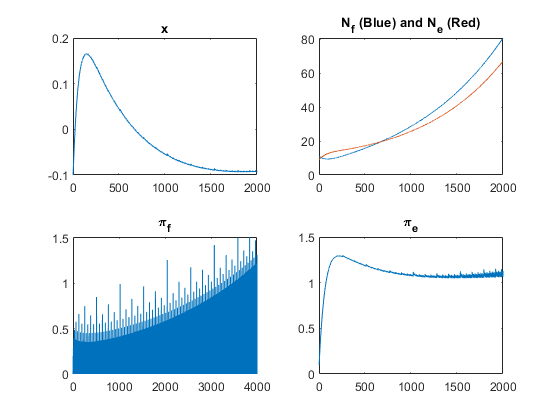}\\
	\caption{Scenario 3: $a_0 = 0.5$ (\emph{Co-existence})}
	\label{fig:sim_3}
\end{figure}

\begin{figure}[h]
	\centering
	\includegraphics[scale=0.8]{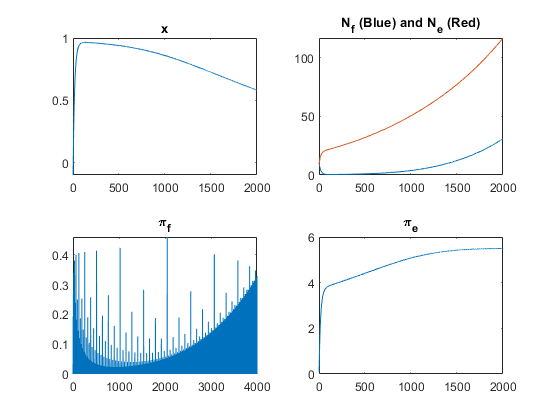}\\
	\caption{Scenario 3: $a_0 = 2.5$ (\emph{Co-existence})}
	\label{fig:sim_4}
\end{figure}

\begin{figure}[h]
	\centering
	\includegraphics[scale=0.8]{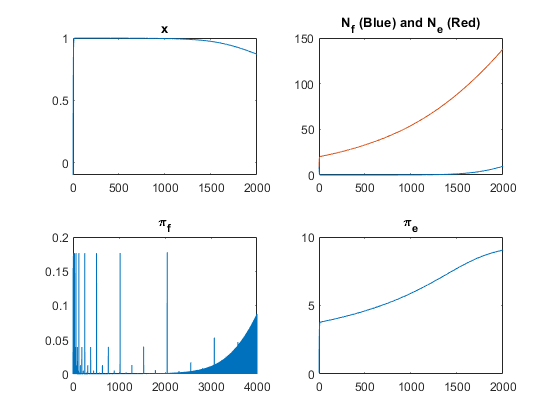}\\
	\caption{Scenario 3: $a_0 = 4.5$ (\emph{Co-existence})}
	\label{fig:sim_5}
\end{figure}

The third scenario looks at the case when both sectors are regulated. We set $a_2=0.5$ and $a_3=-0.5$, which implies that externality from both sectors will be regulated and punished. In this case, the government may still subsidize the NEV sector at an initial stage ($a_0 > 0$), yet the government will also impose corrective tax over NEV-related \emph{Type I} pollution, as soon as $\pi_e$ grows and becomes larger ($a_3=-0.5$). In this scenario, the parameter of initial subsidy will be crucial in determining whether consumers will converge toward TFV or NEV at the end. A low $a_0$ implies that the subsidy is not significant enough, and consumers may eventually choose TFV. Yet a larger $a_0$ may lead to NEV equilibrium in the end. We simulate three cases where $a_0 = 0.5$, $a_0 = 2.5$, $a_0 = 4.5$ respectively, as shown in Fig. \ref{fig:sim_3}-\ref{fig:sim_5}:

\subsubsection{Macro Regulation: Manipulation of $g_N$}
The previous section discusses the long-term dynamics of TFV and NEV when both sectors are monitored and regulated. From the modelling perspective, it is achieved by manipulating $a_2$ and $a_3$. Another aspect of comprehensive regulation may be achieved by imposing the quota on total number of vehicles regardless of TFVs and NEVs. The quota is set by the growth rate of total number of vehicles ($g_N$), which is determined by aggregate externality index $\Pi$:
\begin{eqnarray}
\Pi &=& k_1 \pi_F + k_2 \pi_E
\label{eq:Pi}
\\
g_N &=& \bar{g}_N \exp(-\Pi)
\label{eq:gN}
\end{eqnarray}

Eq. \ref{eq:Pi} defines the aggregate externality index determined by both $\pi_F$ and $\pi_E$. This will in turn determine the cap on $g_N$. If there is no externality, then $g_N = \bar{g}_N$. The presence of externality implies that $g_N$ needs to be lowered, as determined by Eq. \ref{eq:gN}.

\begin{figure}[h]
	\centering
	\includegraphics[scale=0.8]{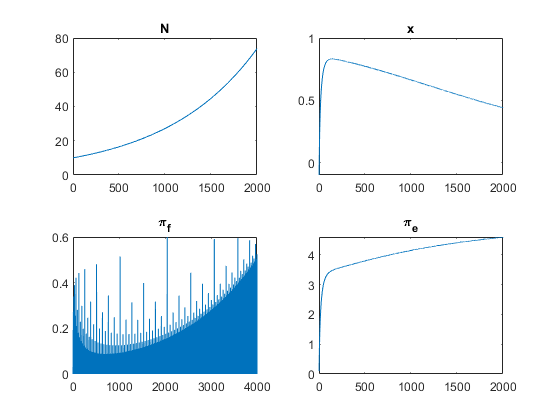}\\
	\caption{$g_N$ grows at an exogenous rate ($\bar{g}_N)$}
	\label{fig:sim_6_1}
\end{figure}

\begin{figure}[h]
	\centering
	\includegraphics[scale=0.8]{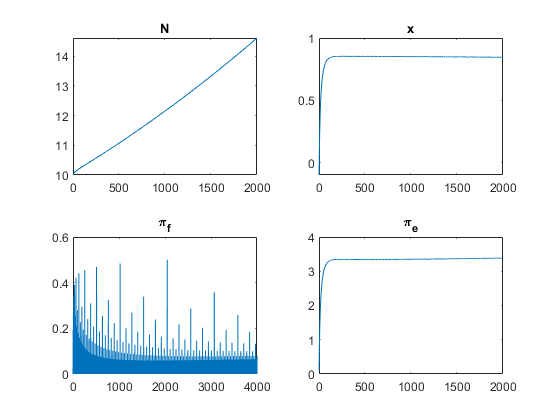}\\
	\caption{$g_N$ under regulation according to $\Pi$}
	\label{fig:sim_6_2}
\end{figure}

The simulation results are shown in Fig. \ref{fig:sim_6_1}-\ref{fig:sim_6_2}. According to this set of simulation, we can see that regulation on $g_N$ can significantly reduce both the number of vehicles ($N$), as well as the associated externality. The real-world implication is that effective regulation requires not only on the micro level in each sector, but also on the macro level on the growth of total number of vehicles. As the government increases spending on public transportation, $g_N$ can be effectively reduced to a desirable level, thus reducing the overall level of environmental externality captured by $\Pi$.

\subsubsection{Other possible scenarios}
The previous sections have discussed the possible long-term dynamics of TFV \& NEV under different policy regimes with an emphasis on parameters in Eq. \ref{eqn_x}. However, there are other parameters in Eq. \ref{eqn_pif}-\ref{eqn_pie} that may deserve further discussion, since they are also subject to policy control. For example, the \emph{R \& D} on fuel efficiency may lower the parameter $\gamma_F$. Similarly, more efficient use of rare-earth elements (REEs) and other related materials such as lithium during NEV production would lower $\theta_E$. Also, new development in circular economy may lead to more efficient recycling of retired TFVs and NEVs, thus increasing the reversing parameters $\alpha_1$ and $\alpha_2$. These non-economic aspects are more difficult to determine and measure, and requires careful cost \& benefit analysis. These aspects are beyond the scope of this paper, which will be investigated further in future research.

\section{Conclusion}
\label{Conclusion}
The ever-worsening global warming and other fuel-related environmental issues have led to an ongoing debate of whether new energy vehicles (NEVs) will eventually ``creatively destroy'' traditional fuel vehicles (TFVs). It is generally believed that NEVs are more efficient, produce zero emission, and will bring a better future to our planet. However, there is a danger that NEV production may also bring unintended environmental consequence due to the use of rare earth elements (REEs) and other pollutants such as lithium, once reaching a large scale of production. Despite its importance, the potential environmental externality arising from NEV production is scantly discussed in current literature.

This paper aims to fill this gap by proposing a simple, preliminary framework that models the long-term dynamics of TFV and NEV growth, as well as their associated environmental externality. We have defined two types of externality in this paper: \emph{Type I} externality that comes from production and \emph{Type II} externality that comes from ongoing usage. We further conduct policy experiments in three aspects: \emph{laissez-faire}, one-sided regulation, and comprehensive regulation. We argue that neither \emph{laissez-faire} nor one-sided regulation would bring us an optimal outcome in the long run, and it is crucial to constantly monitor and regulate the \emph{Type I} externality as well from NEV production at this infant stage of NEV development, so that water and soil pollution is under control. Also, comprehensive regulation implies that TFV and NEV may co-exist for a sustained period of time, so the total externality will not stem from a single source that may bring an irreversible damage to the ecological system. We also suggest that the regulation on the growth of total number of vehicles is more desirable, since it directly reduces the quantity of both TFVs and NEVs, thus significantly lowers total externality. This can be achieved by a long-term commitment of public policies that aims for better infrastructure building and better provision of public transport system, so that the overall environmental impact is kept at its minimum.

There are numerous paths that deserve further investigation. \emph{First}, the model can potentially be calibrated with the use of both economic and environmental data so that we can produce a more realistic long-term projection. \emph{Second}, it is possible to derive a long-term ``golden-rule'' policy recommendation that aims to maximize long-term economic benefit, while minimizing total environmental externality. \emph{Third}, we may look at more realistic consumer behaviour beyond the simple \citet{Lux1995} assumptions. These aspects will be discussed more thoroughly in future research.
%

\section*{Acknowledgement}
\noindent Our work was supported in part by the Guangdong Provincial Key Laboratory of Interdisciplinary Research and Application for Data Science, BNU-HKBU United International College (2022B1212010006) and in part by Guangdong Higher Education Upgrading Plan (2021-2025) (UIC R0400024-21).

\section*{Author Contributions}
\noindent The authors are solely listed in alphabetical order, and all authors have made equal contribution toward this research. Zhaojia Huang derived the analytical results of the 2D and 3D model. Liang Zhang conducted the thorough literature review. Tianhao Zhi ignited the initial idea, constructed the 2D \& 3D model, and worked on the numerical simulations \& policy discussions. All remaining errors are on our own.

\newpage
\bibliographystyle{elsarticle-harv}
\bibliography{Bibliography}

\newpage
\section*{Appendix: Simulation Results}

\end{document}